\documentstyle[11pt,paspconf]{article}

\input{epsf}
\def\be{\begin{equation}}
\def\ee{\end{equation}}
\def\bea{\begin{eqnarray}}
\def\eea{\end{eqnarray}}
\def\etal{\it et al. \rm}

\def\kmsmpc{km~s$^{-1}$~Mpc$^{-1}\,$}

\def\Mo{{\rm M_\odot}}

\def\gsim{ \lower .75ex \hbox{$\sim$} \llap{\raise .27ex \hbox{$>$}} }
\def\lsim{ \lower .75ex \hbox{$\sim$} \llap{\raise .27ex \hbox{$<$}} }
\def\spose#1{\hbox to 0pt{#1\hss}}
\def\simlt{\mathrel{\spose{\lower 3pt\hbox{$\mathchar"218$}}
     \raise 2.0pt\hbox{$\mathchar"13C$}}}
\def\simgt{\mathrel{\spose{\lower 3pt\hbox{$\mathchar"218$}}
     \raise 2.0pt\hbox{$\mathchar"13E$}}}
\def\gsim{ \lower .75ex \hbox{$\sim$} \llap{\raise .27ex \hbox{$>$}} }
\def\lsim{ \lower .75ex \hbox{$\sim$} \llap{\raise .27ex \hbox{$<$}} }
\def\simprop{ \lower .75ex \hbox{$\sim$} \llap{\raise .27ex \hbox{$\propto$}} }
\begin{document}

\title{Numerical and Analytical Modelling of Galaxy Formation and
Evolution} 

\author{C.S. Frenk, C.M. Baugh, S. Cole} 

\affil{Physics Department, University of Durham, Durham DH1 3LE, UK}

\author{C. Lacey} 

\affil{Theoretical Astrophysics Center, DK-2100 Copenhagen 0, Denmark} 


\begin{abstract}
We review recent developments in theoretical studies of galaxy formation
and evolution. In combination with new data from HST, Keck and other large 
telescopes, numerical and semi-analytic modelling is beginning to build up
a coherent picture of galaxy formation. We summarize the current status of
modelling of various galactic properties such as the structure of dark
matter halos, the galaxy luminosity function, the Tully-Fisher relation,
the colour-magnitude relation for ellipticals, the gross morphological
properties of galaxies and the counts of faint galaxies as a function of
magnitude, redshift and morphology. Many of these properties can be
explained, at least at some level, within a broad class of CDM cosmologies,
but a number of fundamental issues remain unresolved. We use our
semi-analytic model of galaxy formation to interpret the evolutionary
status of the Lyman-break galaxies at $z\simeq 3-3.5$ recently discovered by
Steidel \etal The abundance and global properties of these objects are
compatible with model predictions in a variety of CDM cosmologies,
including the standard version. All these models predict mild evolution in
the distribution of star formation rates which peaks at around $z\simeq 1$,
but is never much larger than it is at present. The Steidel \etal
Lyman-break galaxies are among the very first objects in which appreciable
star formation is taking place; they thus signal the onset of galaxy
formation. We present three example evolutionary histories of
Lyman-break galaxies which illustrate that these objects are the precursors
of present day, normal, bright ellipticals and spirals.
\end{abstract}

\section{Introduction}

The spectacular new data obtained in the past year or two with HST, Keck
and other large telescopes are finally giving substance to the adjective
``few'' in Y.B.  Zel'dovich's famous 1977 statement: ``It will only be a
few years before the origin and evolution of galaxies is understood.'' A
full understanding of the origin and evolution of galaxies, however, will
require a great deal of detailed theoretical modelling to uncover the
physical processes manifested in the data. At present such modelling is
lagging behind observational advances partly because of the breathtaking
pace of these advances and partly because some of the physical processes at
work, particularly those involving gas dynamics and star formation, are
intrinsically very complex.

Two interrelated techniques are available for theoretical modelling of
galaxy formation and evolution: numerical simulations and semi-analytic
modelling.  The overall strategy is the same in both cases: to calculate
how density perturbations emerging from the Big Bang turn into visible
galaxies. This requires following a number of processes: (i) the
growth of dark matter halos by accretion and mergers, (ii) the dynamics of
cooling gas, (iii) the transformation of cold gas into stars, (iv) the
spectrophotometric evolution of the resulting stellar populations, (v) the
feedback from star formation and evolution on the properties of prestellar
gas and (vi) the build-up of large galaxies by mergers.  Numerical
simulations so far have focussed on a small subset of these
processes which are treated in as realistic a way as is allowed by current
algorithms and computing power. The semi-analytic approach, on the other
hand, considers the combined effect of all these processes which are
simplified into parametric rules distilled from simulations or analytic
considerations.

The numerical and semi-analytic approaches are clearly complementary and
have different strengths and weaknesses. The simulations generally attempt
to model the relevant physics from first principles, but still require
various approximations and free parameters. For example, when dealing with
gas dynamics one needs to chose between Lagrangian methods like ``Smooth
Particle Hydrodynamics (SPH)'' (Katz \& Gunn 1991, Navarro \& White 1993,
Steinmetz \& Muller 1995, Evrard \etal 1994)
or Eulerian methods (Cen \& Ostriker 1993).
There are also choices to be made regarding the cooling
processes to be included, the mechanism to lay down initial conditions,
the resolution of the calculation, etc. For more realistic modelling of
galaxies, it is also necessary to include ad hoc algorithms for turning
cold gas into stars and for coupling the energy liberated by
stellar winds and supernovae to the gas. 

In the semi-analytic approach (Cole 1991, White \& Frenk 1991, Kauffmann
\etal 1993, Lacey \etal 1993, Cole \etal 1994), the required
approximations and free parameters are more readily apparent. The backbone
of this technique is a Monte-Carlo implementation of the ``extended
Press-Schechter theory'' (Bower 1991, Bond \etal 1991) used to describe the formation
of dark matter halos by hierarchical clustering and merging (Kauffmann \&
White 1993, Lacey \& Cole 1993). An
attractive feature of this approach is that within a fairly general
framework, a full model of galaxy formation is specified by a surprisingly
small number of free parameters. This a common feature of the two main
semi-analytic models in existence today, that of G. Kauffmann and
collaborators (Kauffmann \etal 1993, 1994, Kauffmann 1996, 1996) 
and that of the present authors (Cole \etal 1994, Heyl \etal 1995, Baugh,
Cole \& Frenk 1996a, 1996b).

We summarize here the free parameters that appear in our semi-analytic
model since this will be used in the remainder of this paper.  Within a
given cosmology, the model requires fixing five parameters (see Cole \etal
(1994) for further details): (i) the star formation timescale, i.e. the
timescale on which gas that has cooled inside a dark matter halo is turned
into stars; (ii) a feedback parameter which determines the efficiency with
which energy liberated from supernovae and stellar winds reheats gas
cooling inside a halo; (iii) the initial mass function of the stars that
form; (iv) the timescale on which a galaxy falling onto a halo merges with
the central galaxy and (v) the fraction of the stellar mass in stars above
the hydrogen burning limit. To describe the broad morphology of a galaxy
(i.e. its bulge-to-disk ratio) a sixth parameter is required: (vi) a
threshold mass fraction for a merger to turn a disk into a spheroid.  It
must be emphasized that these are not fitting parameters but rather
parameters that describe various astrophysical processes, mostly related to
star formation, that are poorly understood. Lacking a full physical
understanding of these processes, it seems sensible to adjust the
parameters so as to obtain as good a match as possible to a few basic
observational data, in our case, to the local galaxy luminosity functions
in the B and K bands.

In this article, we summarize some of the lessons learned from
numerical and semi-analytic models of galaxy formation, highlighting a
number of unresolved issues (Section~2). We then deploy our semi-analytic
tools to explore the implications of the recent discovery by Steidel 
\etal (1996) 
of a population of star forming galaxies at redshift $z>3$
(Section~3). We conclude with a brief discussion in Section~4. 

\section{A summary of current theoretical issues} 

Most theoretical work on galaxy formation is carried out within the
framework of hierarchical clustering and gravitational instability (eg
White \& Rees 1978, Peebles 1980). Within this general picture, the
various relevant processes are understood at different levels. Progress in
several of these areas may be summarized as follows.

\medskip

\noindent $\bullet$ {\it Dark matter halos}. 
Processes associated with the gravitational evolution of dark matter halos
are reasonably well understood. This subject has progressed significantly
in the past 15 years as a result of the increased sophistication of N-body
simulations allied to some degree of analytic insight. Thus, in a given
cosmological model, the abundance of dark matter halos, their merging
history and their internal structure can be predicted reliably (Press \&
Schechter 1974, Frenk \etal 1988, Lacey \& Cole 1993, 
Navarro, Frenk \& White 1996, Cole \& Lacey 1996). 
For example, recent high resolution N-body simulations by
Navarro, Frenk \& White (1996)  have established that independently of the
cosmological model, dark matter halos of all masses develop a mass density
profile that follows a simple, two-parameter form, scaling like $r^{-1}$
in the central regions and like $r^{-3}$ near the virial radius. The two
parameters of the fit, which can be expressed as the mass and
characteristic density of each halo, turn out to be strongly correlated:
low-mass halos are significantly denser than more massive halos because,
on average, they form earlier. Thus, in effect, the spherically averaged 
density profiles of dark matter halos are described by a universal 
one-parameter function.

\medskip

\noindent $\bullet$ {\it The shape of the luminosity function}. Both
numerical simulations and analytic considerations indicate that the mass
function of galactic halos has a steeper slope at the low-mass end
($\alpha \simeq 2$) than the observed field galaxy luminosity function
($\alpha \simeq 1$) (Loveday \etal 1992, but see Marzke \etal 1994). The 
semi-analytic models, however, have 
demonstrated that the faint end of the galaxy luminosity function is determined
by the combined effect of mergers and feedback but, in general, no model
so far has succeeded in producing a faint end slope much flatter than
$\alpha \simeq 1.5$. At the bright end, the galaxy luminosity function cuts
off exponentially, much as observed, as a result of the large cooling time
of gas in massive halos.

\medskip 

\noindent $\bullet$ {\it The Tully-Fisher relation}. The Tully-Fisher
relation predicted in semi-analytic models in a variety of cosmologies has
a slope and scatter quite similar to those observed (White \& Frenk 1991, 
Kauffmann \etal 1993, Lacey \etal 1993, Cole \etal 1994, Heyl \etal 1995).
However, so far it has proved impossible to match simultaneously the
zero-point of this relation and the amplitude of the galaxy luminosity
function. The overall luminosity normalization of the models (parameter
(v) above) can be chosen to match one or the other, but not both. This is
another outstanding problem and results from the overabundance of galactic dark
halos predicted in the models. The problem is particularly severe for
standard CDM, but it is still present in low-density variants of this
cosmology.

\medskip 

\noindent $\bullet$ {\it The colours of galaxies}. Standard stellar 
population synthesis models and a standard IMF are sufficient to produce 
model galaxies with the spread of colours observed in the local population.
This is true in most popular cosmologies except in the mixed dark matter 
model in which galaxies are much too young and thus much too blue compared 
to observations (Heyl \etal 1995).

\medskip 

\noindent $\bullet$ {\it The morphologies of galaxies}. N-body/gas dynamic
simulations produce galaxies with spiral disks and bulges (Katz \& Gunn
1991, Steinmetz \& Muller 1995) and merger remnants that resemble
ellipticals (Barnes \& Hernquist 1996, Mihos \& Hernquist 1996). However,
gaseous disks in simulations with realistic initial conditions have much
smaller radii than observed disks because the fragments from which they 
form lose angular momentum to the halo as they merge (Navarro \& Benz 1991,
Navarro, Frenk \& White 
1995). Thus, contrary to common belief, the origin of the angular momentum
of spiral disks is not yet fully understood. Incorporating a simple
prescription for merger-induced transformations of disks into spheroids in
semi-analytic models reproduces the Dressler (1980) morphology-density
relation (Kauffmann 1995, Baugh \etal 1996a). This success provides
suggestive support for the view that accretion of rotating gas and mergers
are the key ingredients in understanding the broad morphological
characteristics of galaxies. The same environmental effects in clusters
that produce the morphology-density relation today are responsible for the
Butcher-Oemler effect (Kauffmann 1996, Baugh \etal 1996a).

\medskip 

\noindent $\bullet$ {\it The colour-magnitude relation of cluster
ellipticals}. Semi-analytic models tend to produce colour-magnitude
relations with an approximately flat slope and small scatter (Kauffmann
1996, Baugh \etal 1996b). This is a counterintuitive result in hierarchical
clustering models in which small objects form first and might therefore 
be expected 
to be redder. It arises because star formation in subgalactic 
fragments generally precedes the assembly of the galaxy by mergers.
Furthermore, elliptical galaxies tend to form from fragments whose mass is
biased towards large values. The traditional argument that the small
scatter in the colour-magnitude relation requires ellipticals to be old
and mergers to be unimportant thus appears to be
incorrect. However, the observed colour-magnitude relation has a small but
non-negligible slope (Bower, Lucey \& Ellis 1992). In the context of current models,
this must arise from metallicity effects which are neglected at
present. It remains a major challenge for the models to reproduce the 
observed slope while retaining a small scatter.

\begin{figure}[ht]
{\epsfxsize=11.5truecm \epsfysize=8.truecm 
\epsfbox[-50 435 590 720]{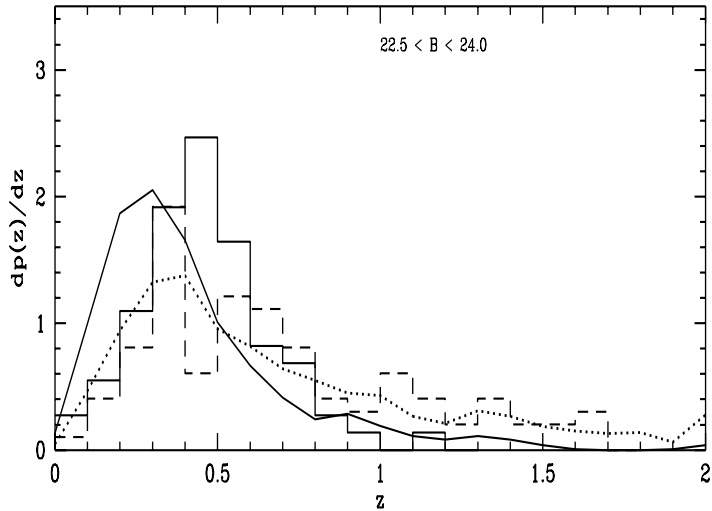}}
\caption[junk]
{
The redshift distribution of galaxies with magnitudes in the range $22.5 <
B < 24.0$. The solid histogram shows the data of Glazebrook {\it et al}
(1995), while the dashed histogram shows the (more complete) data of Cowie {\it et
al} (1996). The lines show the predictions of the model of Cole {\it
et al} (1994) for a Scalo IMF (solid line) and a Miller-Scalo IMF
(dotted line).
}
\label{fig:dndz}
\end{figure}

\medskip 

\noindent $\bullet$ {\it Counts of faint galaxies as a function of
magnitude, redshift and morphology}. A notable success of the semi-analytic
models is the excellent match they provide to the counts of faint galaxies
as a function of magnitude, redshift and morphology. The supporting data
are presented in the papers by White \& Frenk (1991), Lacey \etal (1993), 
Kauffmann \etal (1993), Cole \etal (1994), and Baugh {\it et al} (1996a). 
The agreement is particularly good in the standard CDM
model but it is also acceptable in low-density CDM models.  Particularly
noteworthy is the match to the morphological data from the Hubble Deep
Field discussed by Baugh \etal (1996a) and the prediction that faintwards
of $I\simeq 25$ the galaxy counts should become increasingly dominated by
irregulars. Also noteworthy is the successful prediction of the redshift
distribution of $B\simeq 24$ mag galaxies. The model predictions of Cole 
\etal (1994), published before the observations were made, are compared 
with the
recent data of Cowie \etal (1996) in Figure~1. This agreement is the most
striking indication so far that the models contain some element of truth.
A consequence of these successes is that the models of Baugh \etal (1996a) 
also give a reasonable match to the redshift evolution of the luminosity
function recently measured from the CFRS survey by Lilly \etal (1995)  and
from a combination of surveys by Ellis {\it et al} (1996). 
It should be noted, however, that
the good match to faint data is due, in part, to the steep faint end slope
in the model luminosity function for local field galaxies.

\section{The Lyman break galaxies}

Steidel \etal (1996) have recently discovered a population of star
forming galaxies at redshift $z\simeq 3-3.5$. In the context of the models
discussed here, these galaxies are among the first objects in
which appreciable star formation has taken place. Because of their great 
importance in understanding galaxy formation, we discuss them here in 
some detail.

\begin{figure}[ht]
{\epsfxsize=11.7truecm \epsfysize=11.7truecm 
\epsfbox[-75 140 590 720]{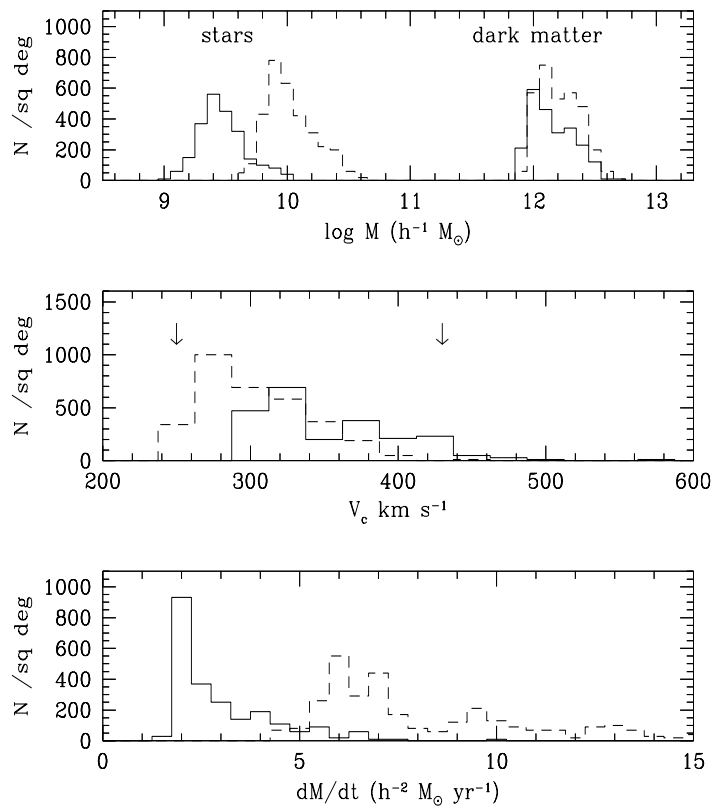}}
\caption[junk]
{
The properties of Lyman-break or ``UV drop-out" galaxies identified in our
models using identical selection criteria to those applied to the
observations of Steidel {\it et al} (1996). Results are given for the
standard CDM model (solid lines) and the $\Lambda$CDM model (dashed
lines). The top panel shows the distribution of stellar mass and halo
mass; the middle panel shows the distribution of halo circular velocities;
and the bottom panel shows the distribution of star formation rates. The 
arrows in the middle panel indicate the range inferred from the 
observations. 
}
\label{fig:steidel}
\end{figure}

Candidate high-z galaxies were identified spectroscopically, using $U_n, G$
and $R$ filters (Steidel, Pettini \& Hamilton 1995).  At $z\simeq 3$ the
912 \AA\ break produced by the Lyman limit shifts into the $U_n$ filter
passband while, for the roughly flat spectrum characteristic of a
star-forming object, the fluxes in the two other filters are
comparable. Follow-up spectroscopy at Keck revealed that the objects so
identified are indeed star-forming galaxies at $3.0\lsim z \lsim
3.5$. Steidel \etal find that these galaxies represent 1.3\% of the faint
counts brighter than $R=25$, corresponding to a comoving number density
comparable to that of present day $L_*$ galaxies.  The spectra of these
galaxies are similar to those of nearby star-forming regions; their
circular velocities are $250\leq V_c/({\rm km\ s}^{-1}) \leq 450$  (if the line widths of
saturated interstellar lines are assumed to reflect the circular velocity
of the galaxy); and their typical star formation rates are inferred to be 
$\sim 2 h^{-2}\Mo$ yr$^{-1}$ for $q_0=0.5$ and $\sim 6 h^{-2}\Mo$ yr$^{-1}$ for 
$q_0=0.05$ (where $h$ is Hubble's constant in units of 100 \kmsmpc.)

At first sight, the existence of a sizeable population of massive
star-forming galaxies at such high redshifts may appear surprising,
particularly in the context of the standard $\Omega=1$, $h=0.5$ CDM
cosmology in which, as has been emphasized for a number of years, galaxy
formation is a relatively recent phenomenon (Frenk \etal 1995). Our
semi-analytic machinery allows us to investigate in detail whether galaxies
with the required properties occur in a given cosmological model. Here we
present results for our standard CDM model (normalised to 
$\sigma_8=0.67$ so as to give the observed local abundance of rich galaxy 
clusters) and for a flat
COBE-normalized ``$\Lambda$CDM'' model ($\Omega=0.3$, $\Lambda=0.7$,
$h=0.6$, $\sigma_8=0.97$). Further details of this analysis and results for
other cosmologies will be presented in a forthcoming paper.

Following our general philosophy, we apply our fully specified model,
i.e. the model in which all free parameters have been previously fixed by
reference to local galaxy data, in essence the model published by Cole {\it
et al} (1994). The only change we have made is to assume a Miller-Scalo
rather than a Scalo IMF. As Cole \etal discuss, the choice of IMF has
little effect on the properties of the local galaxy population but it does
affect the properties of galaxies at high redshift. 
We first selected galaxies using exactly the same 
filters and colour criteria as Steidel {\it et al.}, taking into account the effects of
absorption by intervening cold gas (Madau 1995). These criteria did indeed pick
out galaxies at $2.8\lsim z \lsim 3.5$. The standard CDM model produced
2400 galaxies per square degree brighter than $R=25$ satisfying the colour
criteria, of which 1200 lie in the redshift interval $z=3-3.5$. The
corresponding numbers in the $\Lambda$CDM model are 3700 and 1700. From
their 31 robust candidates, Steidel \etal estimated $1400\pm 300$ galaxies
per square degree in this redshift interval.

The properties of our model ``Lyman-break'' galaxies are displayed in
Figure~2. In the standard CDM model their typical stellar masses are a few
times $10^9 h^{-1}\Mo$ ($\sim 10^{10}h^{-1}\Mo$ in the $\Lambda$CDM model)
and these galaxies inhabit dark matter halos with typical mass $10^{12}
h^{-1}\Mo$. The velocity dispersions of the standard model galaxies are
remarkably similar to those measured by Steidel
\etal Finally, the model star formation rates also agree well with the 
rates inferred by 
Steidel \etal (Star formation rates are not directly measured in the data
but inferred from the $R$ magnitude assuming an IMF and a stellar
population synthesis model which are similar, but not identical, to those
in our galaxy formation model.)

The success of our CDM models of galaxy formation in accounting for the
observed abundance and overall properties of the Steidel \etal galaxies is
both striking and surprising.  However, several caveats are in
order. Firstly, the predicted abundance of star-forming galaxies at high
redshift depends sensitively on at least two model assumptions: the IMF and
the normalisation of the linear density fluctuation spectrum,
$\sigma_8$. Adopting a Scalo rather than a Miller-Scalo IMF reduces the
total number of faint R-band counts by only 20\% but it reduces the number
of Lyman-break galaxies in the redshift interval of interest by about a
factor of 10. Similarly, reducing $\sigma_8$ from our adopted value of 0.67
in the standard CDM model to 0.5 reduces the number of high redshift
galaxies also by a factor of approximately 10. These uncertainties dwarf
the changes produced by varying the other cosmological parameters of the model.

\begin{figure}[ht]
{\epsfxsize=12.truecm \epsfysize=7.truecm 
\epsfbox[-100 435 640 720]{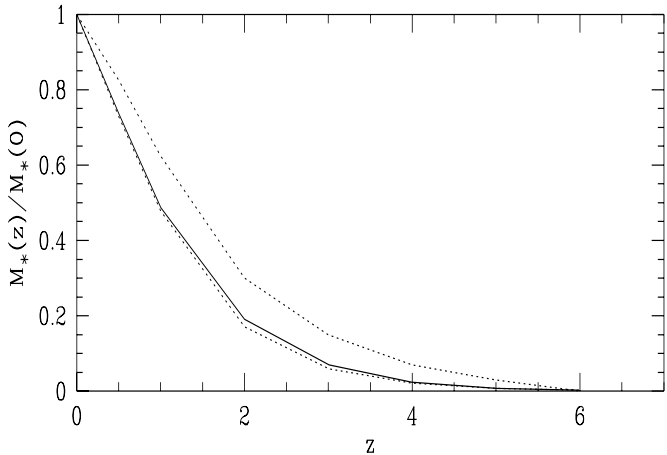}}
\caption[junk]
{
The mass in stars formed by redshift z as a fraction of the final mass in
stars at redshift zero. The solid line shows results for the standard CDM
model, while the dotted lines show results for flat COBE-normalised 
cosmological models 
with $\Omega_0=0.3$, $\Lambda=0.7$ and $h=0.6$. The lower dotted curve has
a spectral shape parameter $\Gamma=0.18$ and the upper dotted curve has
$\Gamma=0.3$. In all cases, less than $10 \% $ of the total mass of stars
has formed by $z=3.5$.
}
\label{fig:mstarsz}
\end{figure}

Regardless of the uncertainties just discussed, the Lyman-break galaxies of
Steidel \etal correspond to the first objects in our models in which
significant star formation is taking place. Figure~3 shows how the overall
stellar population builds up in our two models (and in a variant of the
$\Lambda$CDM model in which the initial power spectrum has more 
power on galactic scales.) In all cases, only a small fraction of the final stellar component
of the Universe has formed by $z=3.5$. The standard CDM and $\Lambda$CDM 
models have almost identical star formation 
histories and both have formed less than 5\% of the total stellar
population by $z=3.5$. In the low-density model with more small scale power
this fraction is still less than 10\%. Thus, in the class of models we are
considering, the redshift $z\simeq 3.5$ at which the Steidel \etal Lyman
break galaxies are found is close to the onset of galaxy formation. Very
few bright objects should exist beyond this redshift.

\begin{figure}[ht]
{\epsfxsize=13.truecm \epsfysize=11.truecm 
\epsfbox[0 150 590 710]{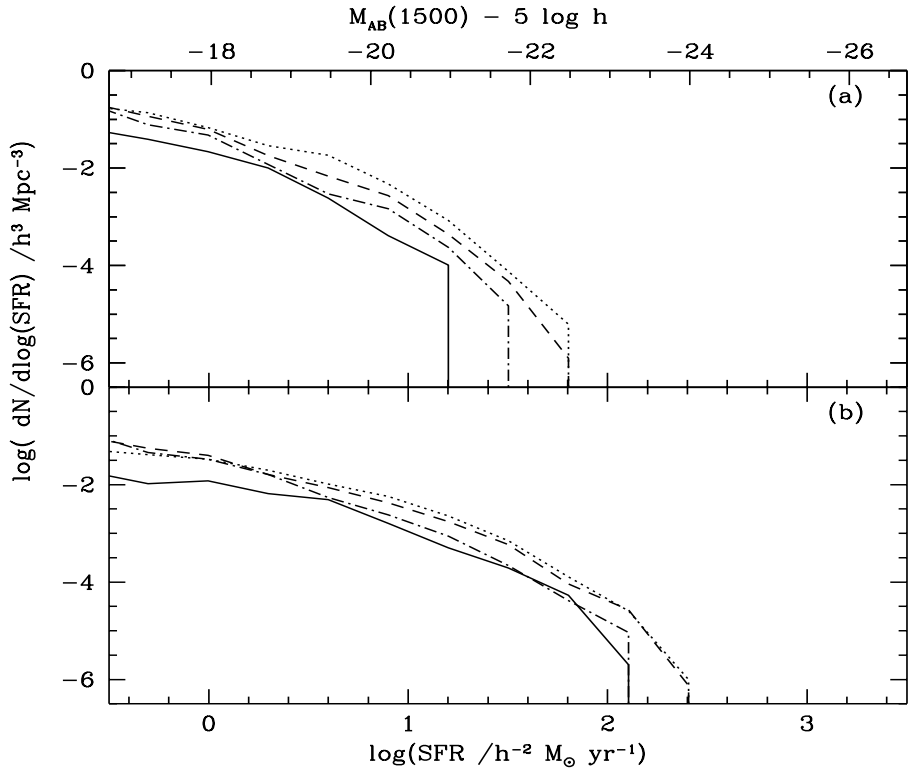}}
\caption[junk]
{
The distribution of star formation rates at four redshifts: $z=3$
(dot-dashed curves), $z=2.35$ (dashed curve), $z=1$ (dotted curves) and
$z=0$ (solid curves). The upper panel shows results for our standard CDM
model and the lower panel for our $\Lambda$CDM model. The distribution of
star formation rates evolves slowly between the epochs shown and is highest
at $z=1$.
}
\label{fig:sfr}
\end{figure}

Our predicted (differential) star formation rates at 
four different redshifts are shown in Figure~4. The upper panel gives
results for the standard CDM model and the lower panel for the
$\Lambda$CDM model. The lower abscissa is labelled by the actual star
formation rate while the corresponding 1500 \AA\ luminosities are given in
the upper abscissa. In both cosmological models, the distribution of star
formation rates has a similar shape at all times, but the rates are higher
at $z=1$ than at $z=3$ or $z=0$. The evolution in the star formation rate
is relatively mild: over most of the range, the comoving abundance of galaxies
varies by less than an order of magnitude between the peak at $z\simeq 1$
and the present.

We can interrogate our galaxy formation model to find out what sort of
objects the Lyman-break galaxies eventually turn into. Two examples taken
from our standard CDM model are shown in the ``tree diagrams" of
Figure~5. Redshift decreases downward in these plots and the width of the
shaded region is proportional to the mass in stars at each epoch. Stars
generally form in subgalactic fragments at high redshift which grow larger
as gas cools onto a disk and turns into stars. Fragments can merge together
and, if the merger is massive enough, the disks turn into a spheroid; a new
disk may grow by subsequent accretion of gas (Kauffmann \etal 1993, 
Baugh \etal 1996b). The galaxy on the 
left of Figure~5 experienced only two very small mergers at $z\simeq 3$ and
grew almost entirely by accretion. This object ends up as a late type
spiral galaxy with a very small bulge. The galaxy on the right formed by
the merger of several fragments, including a major merger at $z\simeq 0.3$.
This galaxy ends up as an elliptical. The asterisks at high $z$ represent
Lyman-break objects that satisfy the Steidel \etal selection
criteria. The spiral galaxy is the descendant of a single fairly massive
Lyman-break object; the elliptical harbours the descendants of two less
massive Lyman-break objects which merged at relatively recent epochs.

The present-day luminosity function of galaxies which had a Lyman-break
progenitor at $3<z<3.5$ is shown in Figure~6 for our two cosmological
models and compared with estimates of the local field luminosity function.
In both models the descendants populate the bright end of the luminosity
function and, in the standard CDM model, most galaxies with $M_B -5 {\rm
log} h \simeq -21$ once harboured a Lyman-break object.

\begin{figure}[ht]
\begin{picture}(300,200)
\put(5,0)
{\epsfxsize=6.3truecm \epsfysize=6.3truecm 
\epsfbox[20 140 590 720]{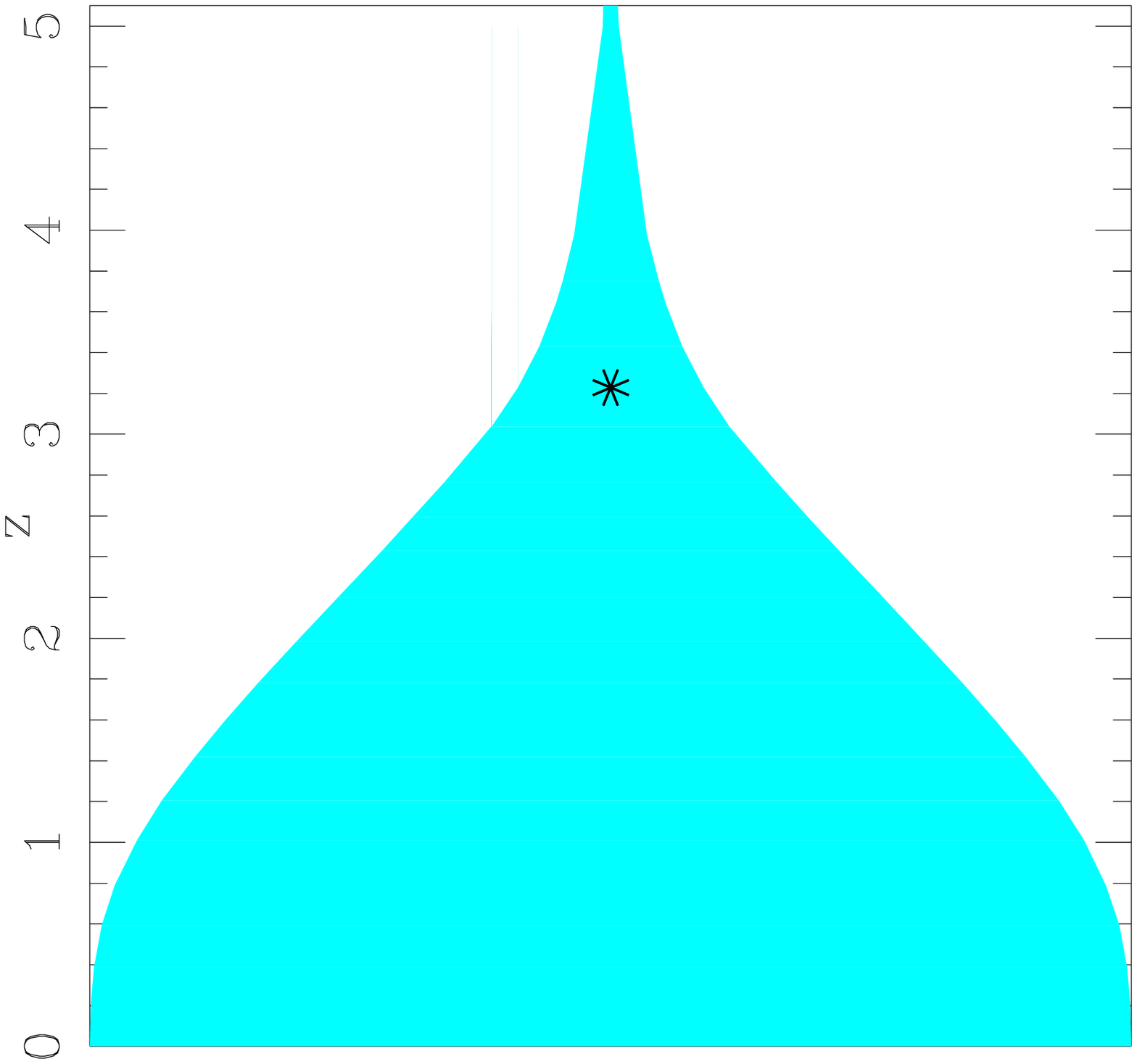}}
\put(190,0)
{\epsfxsize=6.3truecm \epsfysize=6.3truecm 
\epsfbox[20 140 590 720]{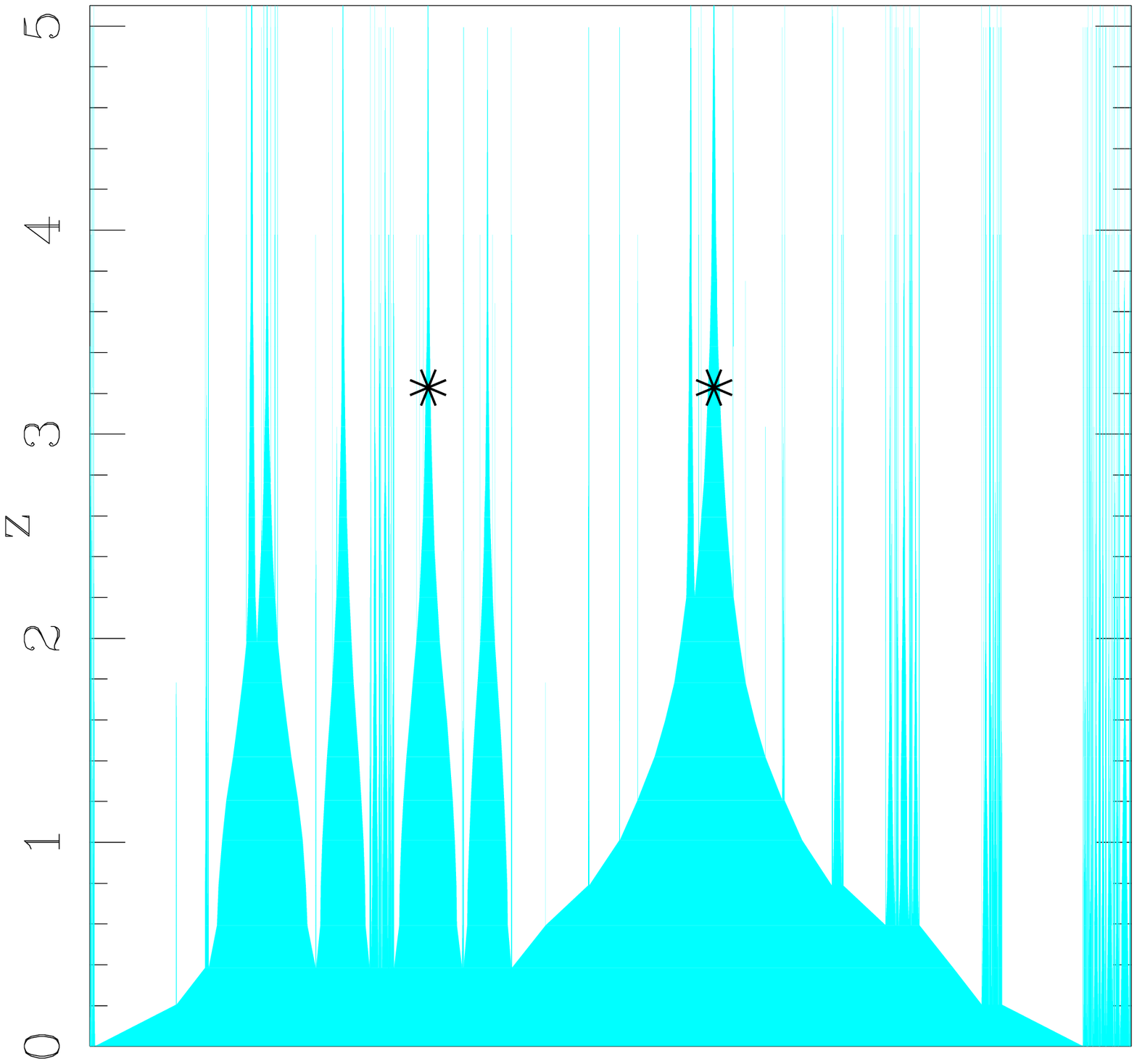}}
\end{picture}
\caption[junk]
{
Star formation histories of two present day galaxies that contained a
Lyman-break progenitor (marked by an asterisk) satisfying the selection
criteria of Steidel {\it et al} (1996). Redshift decreases downward and
the width of the shaded region is proportional to the mass in stars at
each epoch. The galaxy on the left ends up as a late-type spiral; the
galaxy on the right ends up as an elliptical.
}
\label{fig:tree}
\end{figure}
 
\begin{figure}[ht]
{\epsfxsize=12.truecm \epsfysize=7.truecm 
\epsfbox[-80 400 590 720]{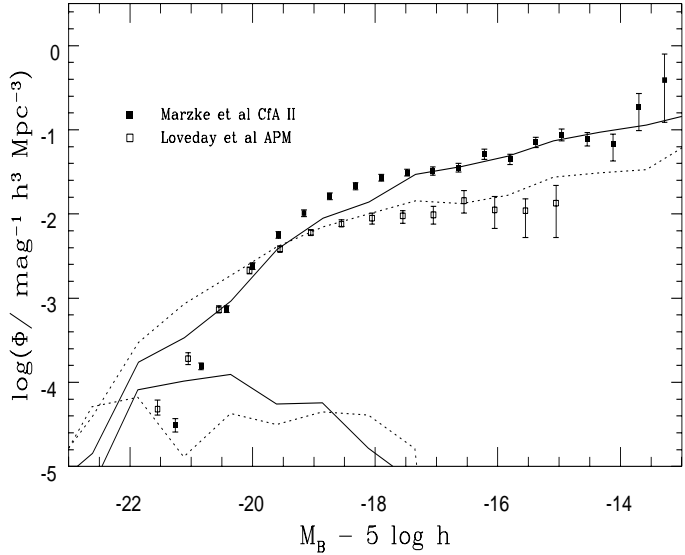}}
\caption[junk]
{
Present-day luminosity functions. The squares show estimates of the field
galaxy luminosity function in the local universe by Loveday \etal (1992;
open symbols) and Marzke \etal (1994; solid symbols). The curves show
predicted galaxy luminosity functions in the standard CDM (solid curves)
and $\Lambda$CDM (dotted curves) models. The curves that extend over the
entire range of magnitudes are the predicted local field galaxy luminosity
functions. The curves near the bottom left of the diagram are the
predicted present-day luminosity functions of galaxies which had a
Lyman-break progenitor at $3<z<3.5$. Many of the brightest galaxies seen
today harboured a Lyman-break object at high redshift.
}
\label{fig:lf}
\end{figure}

\section{Conclusions}

Theoretical studies of galaxy formation, based on numerical simulations and
semi-analytic techniques are an essential complement to observational studies
of the high redshift universe. Such modelling is required in order to 
establish the connection between different types of data and their relation
to the physics of galaxy formation in a cosmological setting. Although
some of the physical processes involved, particularly those associated with
star formation, are very complex and poorly understood, progress can be 
made by complementing a physically based description with heuristic rules 
to describe star formation. 

Semi-analytic models now exist in which the detailed properties of the
galaxy population at all epochs can be predicted {\it ab initio}, starting
from a cosmological spectrum of density fluctuations. The various physical
processes can be characterised by a minimum of free parameters, all of
which are fixed by reference to a small subset of the data for local
galaxies. We have illustrated the predictive power of these semi-analytic
models by comparing our published predictions for the redshift distribution
of galaxies of $B\simeq 24$ mag with recent data from Cowie \etal (1996)
(Figure~1). The excellent agreement between them is the most striking
demonstration so far of the virtues of this approach.

In general, the best understood aspects of galaxy formation are those
related to their dark matter component. The abundance, merging history and
internal structure of galactic halos are all reasonably well established
in a variety of cosmological models of hierarchical clustering. Some
understanding also exists of the physical basis of observable
properties such as the general shape of the galaxy luminosity function,
the slope and scatter of the Tully-Fisher relation, the general features
of the colour-magnitude diagram, the gross morphological properties of
galaxies in different environments, and the counts of faint galaxies as a
function of magnitude, redshift and morphology. All of these properties
can be explained, at least at some level, within a broad class of CDM
cosmologies. 

Several fundamental properties of the galaxy population remain poorly
understood. Examples include the faint end slope of the field luminosity
function which is predicted to be significantly steeper than the standard
estimate (Loveday \etal 1992). None of the existing models can simultaneously
match the zero-point of the Tully-Fisher relation and the overall
amplitude of the galaxy luminosity function, a problem which can be traced
back to an overabundance of dark matter halos predicted in all CDM
cosmologies. While the small scatter in the observed colour-magnitude
relation for cluster ellipticals does not seem incompatible with 
hierarchical clustering, none of the models published to date can
account for the measured slope in this relation.

In spite of the unsolved problems just mentioned, semi-analytic modelling
remains a powerful tool to interpret the recent exciting new data on the high
redshift universe. As an example, we presented
in this article results from new calculations which attempt to identify
the evolutionary status of the Lyman-break galaxies at $z\simeq 3-3.5$
recently discovered by Steidel \etal (1996). Perhaps surprisingly, we
found that the abundance and global properties of these objects is almost
exactly what is predicted by our fiducial model of galaxy formation based
on the standard CDM cosmology. Although the predicted abundance is, in
fact, quite sensitive to certain model assumptions such as the IMF and the
amplitude of mass fluctuations, in general, these galaxies are among the
first objects in which appreciable star formation is taking place. Within
a broad class of CDM models, it appears that the Steidel \etal objects
signal the onset of significant galaxy formation. These objects evolve
into the population of bright normal galaxies seen today. Our models seem
to imply that the long awaited discovery of the early phases of normal
galaxy evolution has now taken place.

\end{document}